# A High-Resolution Radio Continuum Study Of The Dwarf Irregular Galaxy IC 10

J. Westcott,[1]⋆ E. Brinks,[1] R.J. Beswick,[2] V. Heesen,[3,4] M.K. Argo,[2,5] R.D. Baldi,[4] D.M. Fenech,[6] I.M. McHardy,[4] D.J.B. Smith[1] & D.R.A. Williams[4]

[1]*Centre for Astrophysics Research, University of Hertfordshire, College Lane, Hatfield, AL10 9AB, UK*
[2]*Jodrell Bank Centre for Astrophysics, School of Physics and Astronomy, The University of Manchester, Manchester, M13 9PL, UK*
[3]*Hamburger Sternwarte, Universität Hamburg, Gojenbergsweg 112, 21029 Hamburg, Germany*
[4]*School of Physics and Astronomy, University of Southampton, Southampton SO17 1BJ, UK*
[5]*Jeremiah Horrocks Institute, University of Central Lancashire, Preston PR1 2HE, UK*
[6]*Department of Physics & Astronomy, University College London, Gower Street, London, WC1E 6BT, UK*

**ABSTRACT**

We present high-resolution e–MERLIN radio continuum maps of the Dwarf Irregular galaxy IC 10 at 1.5 GHz and 5 GHz. We detect 11 compact sources at 1.5 GHz, 5 of which have complementary detections at 5 GHz. We classify 3 extended sources as compact H II regions within IC 10, 5 sources as contaminating background galaxies and identify 3 sources which require additional observations to classify. We do not expect that any of these 3 sources are Supernova Remnants as they will likely be resolved out at the assumed distance of IC 10 (0.7 Mpc). We correct integrated flux densities of IC 10 from the literature for contamination by unrelated background sources and obtain updated flux density measurements of $354 \pm 11$ mJy at 1.5 GHz and $199 \pm 9$ mJy at 4.85 GHz. The background contamination does not contribute significantly to the overall radio emission from IC 10, so previous analysis concerning its integrated radio properties remain valid.

## 1 INTRODUCTION

Radio emission within normal galaxies originates from the evolution and death of massive stars ($\geq 8 M_\odot$) and hence effectively traces recent star formation (SF; Condon 1992). Since it is not subject to extinction by dust, radio emission should be an ideal tracer of SF which would be valid out to the highest redshifts. At the time of writing, radio luminosity is calibrated as a SFR indicator via the radio–FIR relation (Bell 2003), a tight correlation which links both FIR and radio emission to massive SF. Yet the physical origin of this correlation is not fully understood, with many seemingly independent factors balancing out to form a 'cosmic conspiracy' (Condon 1992; Appleton et al. 2004; Lacki et al. 2010). This then makes it desirable to calibrate the radio–SF relation without recourse to the FIR data.

The **L**egacy **e**–**M**ERLIN **M**ulti–band **I**maging of **N**earby **G**alaxie**s** (LeMMINGs; Beswick et al. 2014) survey aims, in part, to address this by studying compact SF products and their relation to large scale radio emission across a wide range of galaxy type. The project is made up of two complementary surveys of nearby galaxies; a statistical survey of 280 galaxies taken in snapshot mode, comprising the entire Palomar spectroscopic bright galaxy sample north of declination +20 degrees, and a deep survey of 6 galaxies, the latter reaching an rms sensitivity of 8 and 3 µJy beam$^{-1}$ at 1.5 GHz and 5 GHz respectively. The statistical survey will detect and resolve radio emission from supernovae (SNe) and supernova remnants (SNR) to provide a complete census of energetic SF products in the local Universe. At higher sensitivities, the deep survey will also be able to detect faint radio emission from Planetary Nebulae, H II regions and superstar clusters. Together, these surveys will be used to calibrate the SF rate (SFR) in nearby galaxies on the basis of compact radio source populations, independent of obscuration by dust. The nearby dwarf irregular galaxy IC 10 (see Figure 1) was observed during the commissioning phase of the upgraded e–MERLIN array and is the first galaxy to be observed as part of the LeMMINGs deep survey.

⋆ E-mail: j.westcott3@herts.ac.uk

Table 1. Key IC 10 properties

| Property | Value | Reference |
|---|---|---|
| $\alpha_{J2000}$ | 00 20 17.34 | – |
| $\delta_{J2000}$ | +59 18 13.6 | – |
| Galaxy Type | IBm | 1 |
| Distance | 0.7 Mpc | 2 |
| $SFR_{H\alpha}$ | $0.039 \pm 0.001\,M_\odot\,yr^{-1}$ | 3 |
| Angular Size (Major Axis) | $11'.68$ | 4 |
| Angular Size (Minor Axis) | $7'.12$ | 4 |
| Position Angle | $132°$ | 4 |

**Reference List**: 1: Nilson (1973), 2: Sakai et al. (1999), 3: Hunter et al. (2012), 4: Jarrett et al. (2003).

Table 2. Journal of IC 10 e–MERLIN observations

| Obs Date (yyyy mmm dd) | Obs Frequency (MHz) | Total Bandwidth (MHz) | Time (h) |
|---|---|---|---|
| 2013 Feb 09 | 1510.65 | 512 | 7.64 |
| 2013 Nov 22 | 1526.65 | 512 | 16.24 |
| 2015 Apr 18 | 5072.50 | 512 | 21.06 |
| 2016 Feb 21 | 5072.50 | 512 | 9.24 |

**Notes**: Obs Frequency is the central observed frequency and Time is the total time on source.

IC 10 is thought to be currently undergoing a starburst phase (Leroy et al. 2006), with optical studies revealing numerous H II regions (Hodge & Lee 1990) and the highest density of WR–stars observed in the Local Group (Massey et al. 1992; Crowther et al. 2003). Moreover, stellar population analysis reveals a bursty SF history with the most recent burst finishing $\sim10\,Myr$ ago (Hunter 2001; Vacca et al. 2007; Sanna et al. 2009). A brief summary of the properties of IC 10 is presented in Table 1.

Recent radio investigations into the large–scale gas motions of IC 10 reveal a complex nature, consisting of a large H I envelope with an inner rotating disk and counter–rotating component (Shostak & Skillman 1989; Wilcots & Miller 1998; Ashley et al. 2014). The leading explanation for these different kinematical subsystems is that IC 10 is still accreting primordial gas (Wilcots & Miller 1998), although new evidence hints towards a recent merger or interaction being responsible for the unusual gas kinematics and current starburst (Nidever et al. 2013; Ashley et al. 2014). Furthermore, numerous bubbles and shells are observed in the large scale H I distribution (Shostak & Skillman 1989; Wilcots & Miller 1998), hinting that IC 10 is coming towards the end of its current starburst phase (see the bottom–right panel in Figure 1). A number of radio continuum studies focused on the non–thermal superbubble towards the South–East of IC 10 (Yang & Skillman 1993; Heesen et al. 2015), which is thought to be the result of either a collection of supernova remnants (Yang & Skillman 1993) or a single hypernova event (Lozinskaya & Moiseev 2007). At its centre resides the X–ray binary system, IC 10 X–1, which is likely related to the birth and evolution of the superbubble (Bauer & Brandt 2004; Brandt et al. 1997).

IC 10 is located at a low galactic latitude ($\sim-3°$), making distance measurements challenging with estimates ranging from 0.5 Mpc to 3 Mpc (Demers et al. 2004; Kim et al. 2009). In this paper, we will assume a distance to IC 10 of 0.7 Mpc (1 arcsec $\simeq 3.4$ pc), as listed in Hunter et al. (2012); we scale all literature measurements used here to this distance.

The paper is structured as follows; in Section 2 we describe our observations, data reduction and imaging methods; in Section 3 we describe and analyse our source detection method and present our results; in Section 4 we describe the detected sources and discuss our analysis and we summarize our results in Section 5.

## 2 OBSERVATIONS AND DATA REDUCTION

Observations of IC 10 were made using e–MERLIN (the upgraded **M**ulti–**E**lement **R**adio **L**inked **I**nterferometer **N**etwork, an array consisting of 7 radio telescopes spread across the **UK**[1]) as part of LeMMINGs (Beswick et al. 2014), an e–MERLIN legacy project. The data were taken at 2 wavebands; 1.5 GHz (L–Band) and 5 GHz (C–Band) in 4 observing runs summarized in Table 2. The Lovell Telescope was only available for the February 2016 observation, resulting in a smaller field of view but higher sensitivity when compared to the other observing runs. The calibration and data reduction was carried out using AIPS[2], following the calibration steps detailed in the e–MERLIN cookbook[3].

### 2.1 Initial Flagging & Inspection

Before any calibration took place, we used SERPent (Peck & Fenech 2013) with default settings to remove any obvious radio frequency interference (RFI). SERPent is auto–flagging software developed in ParselTongue (Kettenis et al. 2006) principally for use in the e–MERLIN pipeline (Argo 2015). We also conducted further manual flagging to remove other obvious RFI that SERPent had missed. The visibilities from each dataset were then visually inspected using the AIPS task IBLED to further identify and flag time periods when the data was corrupted. We estimate that overall, we are flagging $\sim 52\%$ of the data at 1.5 GHz and $\sim 10\%$ at 5 GHz.

### 2.2 1.5 GHz Data

Both 1.5 GHz observations followed the same observational set–up, which we describe here. The total bandwidth was split into 8 IFs (intermediate frequencies), each with 128 individual channels of bandwidth 0.5 MHz. The primary flux calibrator, 3C286, was used to set the flux scale, the point–like source, OQ208, was used to determine the passbands and relative antenna gains and 0027+5958 was used as the phase calibrator. The observations started with 30 min scans of both the primary flux calibrator and point source, and then went on to alternate between observations of IC 10 and the phase calibrator, spending 5 min on IC 10 and 1 min on the phase calibrator each iteration. We followed standard calibration procedure to determine the complex gain solutions to apply to the data. The first and last 15 channels in each IF were flagged because they are in the least sensitive areas of the passband and contribute mostly noise. Several rounds of phase only self–calibration were carried out on the phase reference source and on IC 10 itself to improve the quality of the final maps. Each calibrated dataset was weighted to give more weight to telescopes with a larger collecting area, improving sensitivity.

---

[1] A basic overview of the array can be found at http://www.e-merlin.ac.uk/summary.html
[2] AIPS, the Astronomical Image Processing Software, is free software available from the NRAO.
[3] Available at: http://www.e-merlin.ac.uk/data_red/tools/e-merlin-cookbook_V3.0_Feb2015.pdf

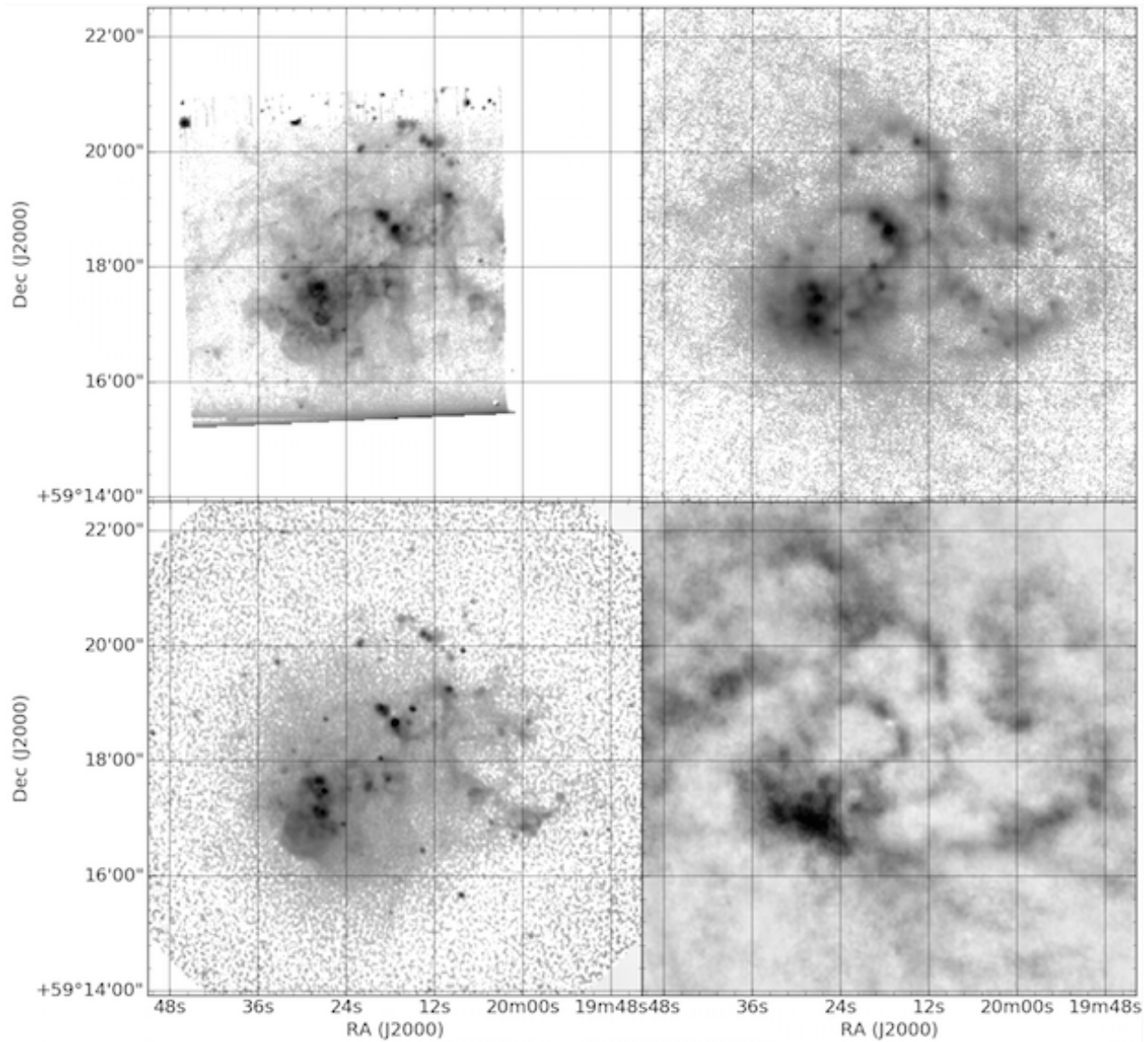

**Figure 1.** IC 10 at multiple wavelengths. *Top Left:* Hα continuum subtracted map taken from Hunter & Elmegreen (2004). *Top Right:* Map of IC 10 at 70 μm taken with Herschel as part of the Dwarf Galaxy Survey (Madden et al. 2013). *Bottom Left:* 5 GHz VLA C & D array map of IC 10 taken from Heesen et al. (2015). *Bottom Right:* Naturally weighted H I map of IC 10 taken as part of LITTLE THINGS (Hunter et al. 2012). Note how the large-scale holes and shells in the H I distribution can be easily identified across multiple wavelengths.

The Cambridge antenna was not available for the November 2013 observation, resulting in a slightly larger synthesized beam as this antenna makes up the longest baselines in the e–MERLIN array. Additionally, IF 1 was flagged because no reliable flux density measurement for OQ208 could be obtained due to excessive RFI.

Both 1.5 GHz observations were calibrated independently and then combined. We verified that both datasets had the same flux scale by imaging 8 bright sources within the primary beam but outside IC 10 (identified in preliminary imaging) and checking that the flux density measurements agreed to within 10 percent error. The flux densities agreed for 7 sources, showing that our flux scale is satisfactorily consistent between the observations. We conclude that the outlier must be down to variability between the observation dates. As the November dataset was taken at a slightly higher frequency than the February dataset, we used the AIPS task BLOAT to ensure that both observations covered the same frequency range before combining both datasets using the task DBCON. The combined *uv*–plane is displayed in Figure 2(a).

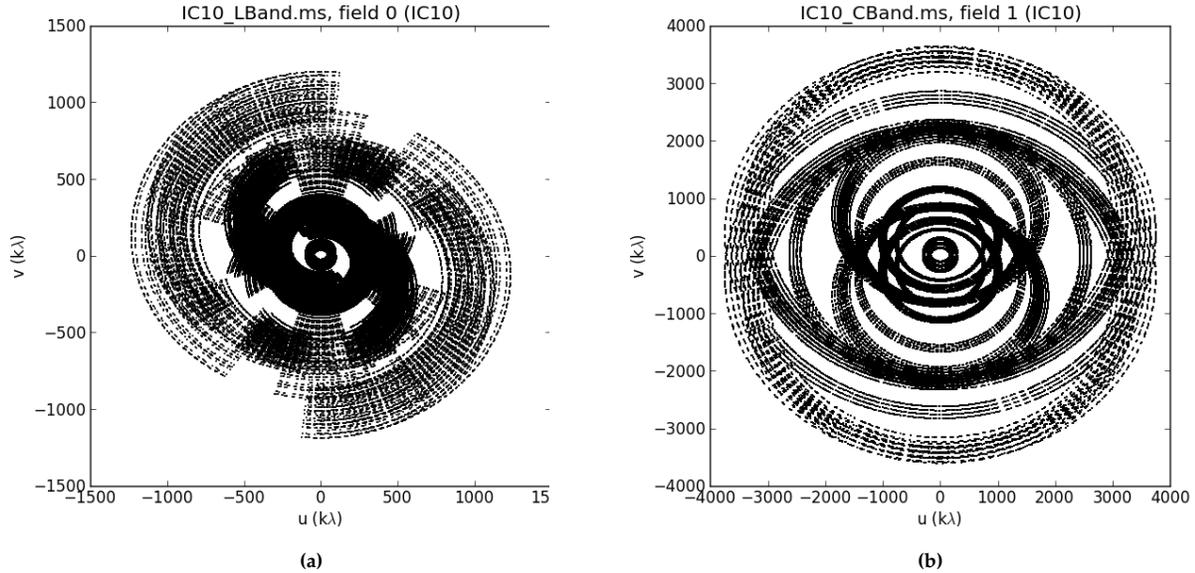

**Figure 2.** Distributions of visibillities in the *uv*–planes for the e–MERLIN observations. **(a)** shows the combined 1.5 GHz *uv*–plane and **(b)** shows the combined 5 GHz *uv*–plane.

### 2.3  5 GHz Data

The 5 GHz observations followed a similar observing set up to the 1.5 GHz observations. The total bandwidth was split into 4 IFs, each with 128 individual channels 1 MHz wide. The primary flux calibrator, 3C286, was used to set the flux scale, the point source, OQ208, was used to determine the passbands and relative antenna gains and 0027+5958 was used as the phase calibrator. IC 10 and the phase reference were observed alternatingly, with 5 min on IC 10 to 3 min on the phase calibrator, finishing with 30 min scans of both the primary flux and point source calibrators. The first and last 5 channels were flagged because this is where the passband rapidly loses sensitivity. One round of phase only self–calibration on IC 10 was performed to further improve the dynamic range of the final maps. Again, both calibrated datasets were weighted.

There were several issues with the February 2016 observation. The Mk2 antenna was not available for the duration of this observation and reliable gain solutions could not be found for the Darnhall antenna. The remaining data were combined with data taken in Apr 2015 to further reduce the noise level in the final maps. The combined 5 GHz uv-plane is displayed in Figure 2(b).

### 2.4  Imaging

IC 10's star forming disk was imaged at 1.5 GHz with a moasic of 19 subfields, with each subfield covering $0.'3$ a side. The AIPS task SETFC was used to generate the list of fields to be imaged and a natural weighting scheme was used to maximise sensitivity. A small field centred on the bright source NVSS002108+591132 was added to this list to remove its contaminating sidelobes. The individual subfields were cleaned down to 2.5 times the rms noise level of each map and were then mosaicked together using the AIPS task FLATN to produce a single widefield map of the entire disk. This widefield map was used to search for sources (see section 3.1) and blank subfields with no apparent sources were used to test the widefield maps completeness (see section 3.6). We image the detected sources separately at 5 GHz to save time, also using a natural weighting scheme to maximise sensitivity. We correct each detected source for the primary beam individually using the models described in section 3.2. This was done to save time as each detected source was suffciently small to be characterised by a single primary beam correction factor.

Our final maps reach a noise level of 26 µJy beam$^{-1}$ at 1.5 GHz and 12 µJy beam$^{-1}$ at 5.0 GHz. These noise levels are significantly higher than the LeMMINGs target sensitivies of 8 µJy beam$^{-1}$ and 3 µJy beam$^{-1}$ at 1.5 GHz and 5 GHz respectively. This can mainly be attributed to the lack of the Lovell telescope in the majority of the presented observations as its inclusion results in a factor of 2 improvement in sensitivity (see e–MERLIN technical capabilities). Additionally, as the array was still in the comissioning phase for most of the observations, some antennas were not present for the entire duration of an observation. It is also likely that the increased noise level is due to the considerable flagging carried out (see Section 2.1) and that low levels of unflagged RFI further contribute to the noise level in the final maps. Together, these factors largely explain the higher noise level. The sensitivity targets are likely to be hit in future LeMMINGs deep observations provided the Lovell telescope will be included for the entire duration of these observations and the e–MERLIN array will be operating at full capacity.

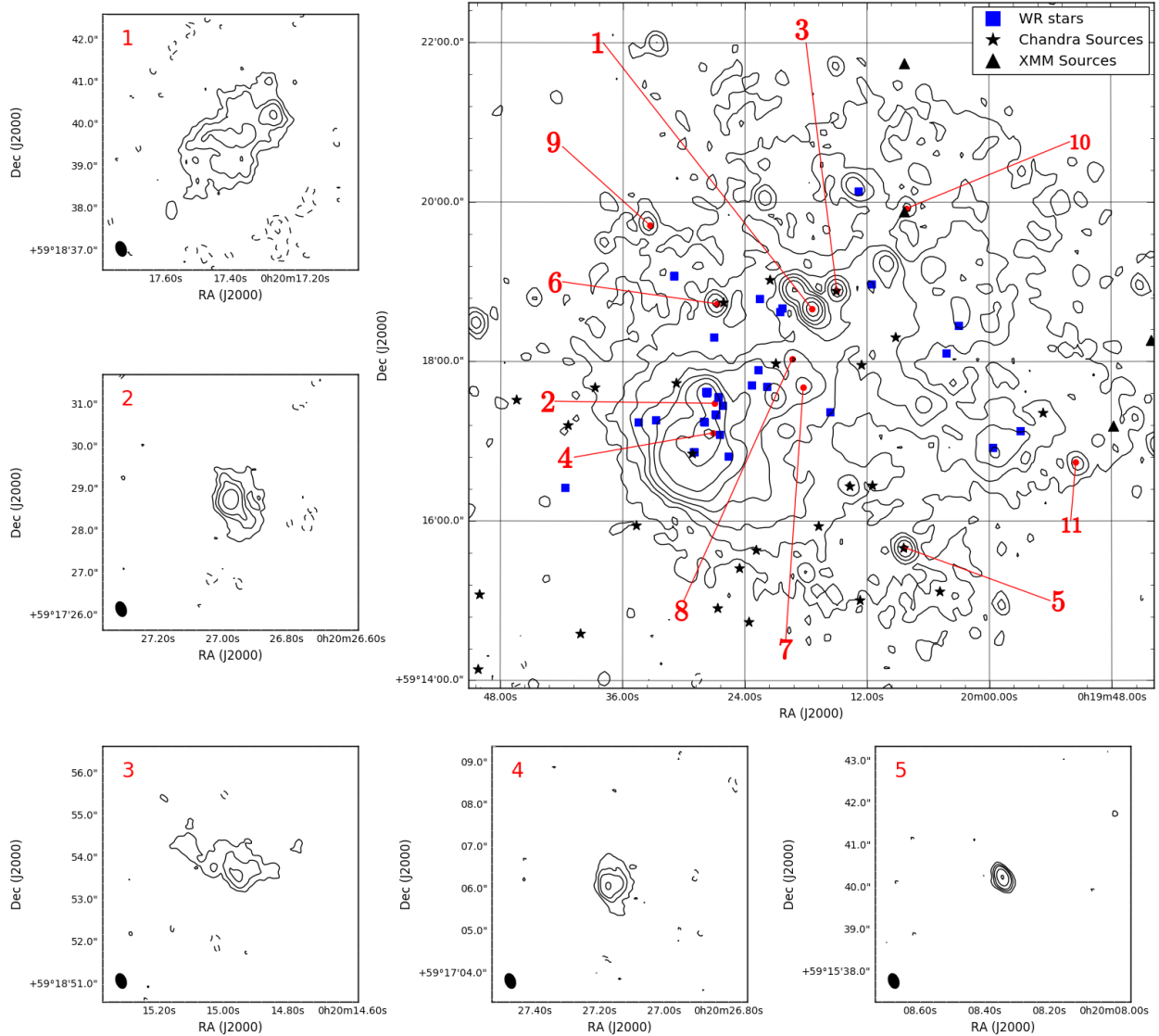

**Figure 3.** 1.5 GHz Radio contours of IC 10. *Top Right*: Combined VLA C+D array map of IC 10 at 1.5 GHz, taken from Heesen et al. (2015). The contours are set to -3, 5, 10, 20, 40, 80, 150, 300, 600 x the 13 µJy beam$^{-1}$ noise level. The red numbers represent the locations of the detected e–MERLIN sources, the blue squares represent spectroscopically confirmed WR stars from Crowther et al. (2003), the black stars represent Chandra (resolution 3″) point source detections and the black upwards triangles represent XMM (resolution 13″) point source detections. All X–ray data are taken from Wang et al. (2005). The gridlines are the same as in Figure 1. *Surrounding plots*: Naturally weighted 1.5 GHz e–MERLIN detections from this study. The beam size for each image is 0.″36 x 0.″24 and is displayed in the bottom-left of each map. The contours on each source are set to -3, 3, 6, 10, 20, 40 x the 26 µJy beam$^{-1}$ noise level. The contours also correspond to brightness temperatures set at 690, 1400, 2300, 4600 and 9200 K. Each field measures 6″ a side. An identifying source number is located in the top–left corner of each contour map.

## 3 RESULTS

Radio emission from star–forming galaxies originates from two main sources, thermal emission from H II regions and non–thermal emission from Cosmic Ray electrons (CRe) that have been accelerated in SNR shock fronts. It is unlikely that any detections are related to X–ray binaries or Planetary Nebulae (Fender & Hendry 2000; Leverenz 2016) as at the distance of IC 10, these sources would have flux densties of order 1 µJy. Very long baseline interferometers, such as e–MERLIN, effectively resolve out nearly all extended emission within observed galaxies, essentially highlighting compact star–forming products and contaminating background sources. In this section we present our e–MERLIN detections within IC 10 and test the robustness of our detections.

### 3.1 Source Detection

We used the widefield maps described in Section 2.4 to search for compact radio sources. We use the map at 1.5 GHz to search for sources because not only does this map have

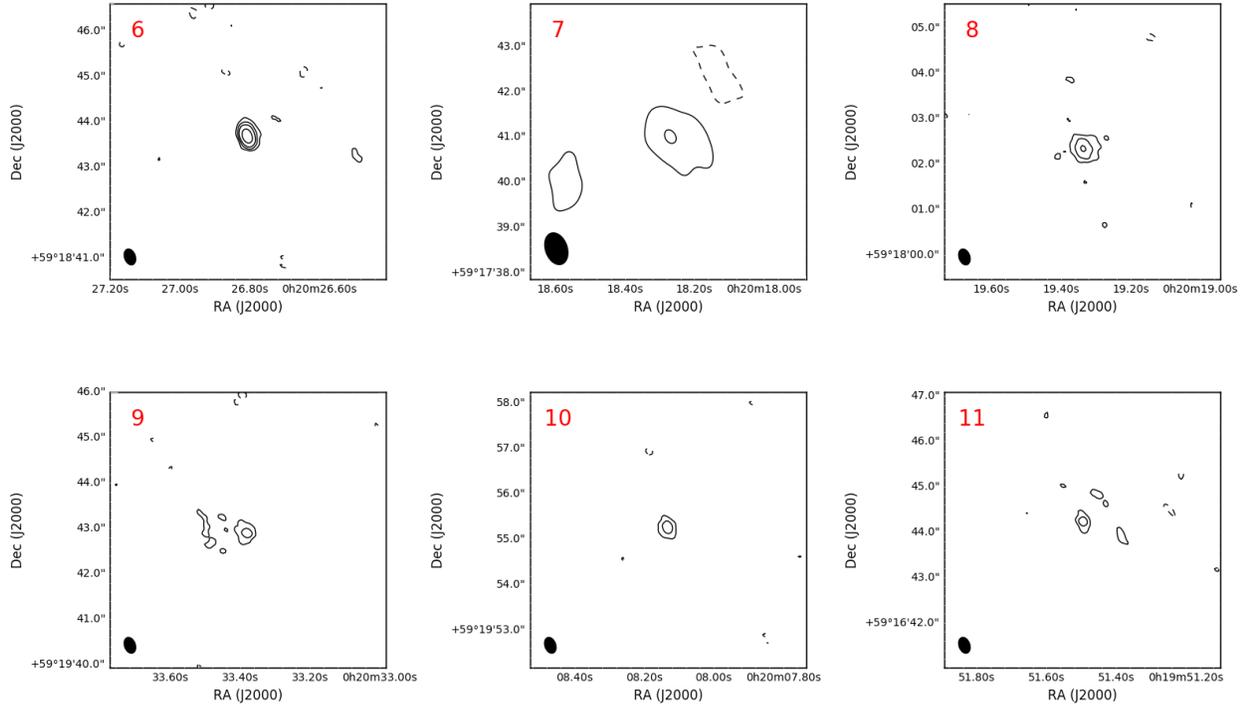

**Figure 3. (Continued.)** *Additional Notes*: The map of Source 7 has been convolved to a resolution of twice the synthesized beam to show the low surface brightness emission. The contours on source 7 are set to the same levels as in the other contour maps, but the noise level in this case is 53 µJy beam$^{-1}$. The contours correspond to brightness temperatures set at 170 K and 340 K in this case only.

**Table 3.** Properties of Sources Detected in IC 10

| Source No | IAU Designation | $\alpha_{J2000}$ (hh mm ss) | $\delta_{J2000}$ (dd mm ss) | $S_{1.5\,GHz}$ (mJy) | $S_{5.0\,GHz}$ (mJy) | Peak $T_B$ (K) | Spectral Index |
|---|---|---|---|---|---|---|---|
| 1* | WBBH J002017+591839.7 | 00 20 17.390 | +59 18 39.68 | 8.06 ± 0.82 | 1.02 ± 0.11 | 5200 | *−0.3 ± 0.1* |
| 2* | WBBH J002027+591728.7 | 00 20 26.952 | +59 17 28.66 | 3.39 ± 0.35 | 2.28 ± 0.24 | 7700 | *−0.1 ± 0.1* |
| 3* | WBBH J002015+591853.9 | 00 20 14.988 | +59 18 53.88 | 2.23 ± 0.25 | - | 2600 | *−1.0 ± 0.1* |
| 4* | WBBH J002027+591706.1 | 00 20 27.160 | +59 17 06.12 | 1.83 ± 0.20 | 0.36 ± 0.06 | 4900 | *−0.4 ± 0.1* |
| 5 | WBBH J002008+591540.2 | 00 20 08.348 | +59 15 40.21 | 1.29 ± 0.14 | 0.28 ± 0.04 | > 11000 | −1.3 ± 0.2 |
| 6 | WBBH J002027+591843.7 | 00 20 26.807 | +59 18 43.69 | 0.99 ± 0.11 | - | > 8500 | ≤ −1.9 |
| 7* | WBBH J002018+591740.9 | 00 20 18.249 | +59 17 40.92 | 0.93 ± 0.14 | - | 700 | *−0.4 ± 0.1* |
| 8 | WBBH J002019+591802.3 | 00 20 19.334 | +59 18 02.33 | 0.87 ± 0.16 | - | > 7700 | ≤ −2.0 |
| 9 | WBBH J002033+591942.9 | 00 20 33.390 | +59 19 42.89 | 0.49 ± 0.11 | - | > 4100 | ≤ −1.0 |
| 10 | WBBH J002008+591955.2 | 00 20 08.137 | +59 19 55.23 | 0.37 ± 0.07 | 0.14 ± 0.03 | > 3200 | −0.8 ± 0.3 |
| 11 | WBBH J001951+591644.2 | 00 19 51.498 | +59 16 44.20 | 0.26 ± 0.06 | - | > 2000 | ≤ −1.1 |

**Notes**: Sources with a * are extended sources and their spectral indices (italicized) are derived from the low–resolution spectral index maps presented in Heesen et al. (2011). All other spectral indices are derived from the e–MERLIN observations. The peak brightness temperatures have been derived from the 1.5 GHz observations.

a wider field of view, it is more sensitive to extended emission than the corresponding 5 GHz maps (see Figure 2). Furthermore, any sources dominated by non–thermal emission mechanisms will be brighter at 1.5 GHz due to the power–law radio spectra these sources exhibit. We used Aegean as our source extractor (Hancock et al. 2012) with a peak flux density threshold of 5 times the rms noise level, $\sigma_{rms}$, to find candidates. The source extractor retrieved 26 sources in total, of which 15 are coincident with IC 10's main disk. We cross-matched these remaining candidates with lower resolution features from Karl G. Jansky Very Large Array (VLA) maps at 1.5 GHz (Heesen et al. 2015; see Figure 3) to identify spurious detections. The lower resolution maps have a 13 µJy beam$^{-1}$ rms noise level, and so, we would expect to observe a 5 $\sigma$ e–MERLIN detection in these maps. We determine 2 of the e–MERLIN detections to be spurious and end up with a sample of 13 sources which we display in Figure

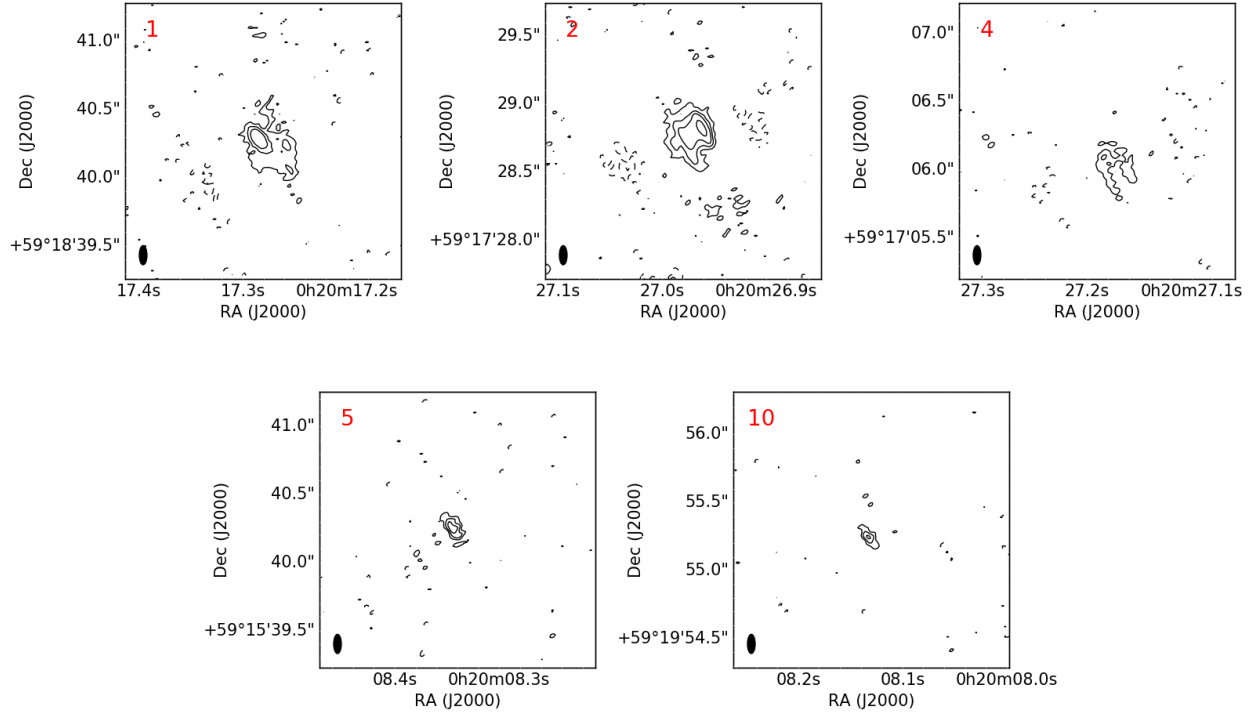

**Figure 4.** Naturally weighted 5 GHz e–MERLIN detections from this study. The beam size for this image is $0''.14 \times 0''.06$ and is displayed in the bottom–left of each contour plot. The contours on each source are set to -3, 3, 6, 10, 20, 40 x the 12 μJy beam$^{-1}$ noise level. Each field measures $2'$ on a side. An identifying source number is located in the top-left corner of each contour map.

3. We discuss the completeness and reliability of our source detection method in Section 3.6.

### 3.2 Source Flux Measurement

We determine integrated flux densitites for the unresolved and resolved sources using 2 different methods. For unresolved sources we fit Gaussians through the use of the AIPS task JMFIT. For resolved sources, we generate a mask by only considering emission associated with each source above a $3\,\sigma_{\rm rms}$ level, where $\sigma_{\rm rms}$ is the local map rms noise level (measured off source). We then integrate the masked emission to find the integrated flux density.

The errors for the integrated flux density measurements are determined using:

$$\sigma_{\rm Total} = \sqrt{(\sigma_{\rm Map})^2 + (\sigma_{\rm Scale})^2} \qquad (1)$$

where $\sigma_{\rm Map}$ is either equal to the fit error returned by JMFIT for unresolved sources or it is equal to $\sqrt{N}\sigma_{\rm rms}$ for resolved sources ($N$ is the number of independent beams that fit within the masked image). $\sigma_{\rm Scale}$ is the uncertainty in the flux scale which is taken to be 10 per cent.

We correct the integrated flux density from each source for primary beam attenuation individually by assuming the e-MERLIN primary beam can be parameterized by a Gaussian, as displayed in the e–MERLIN techincal capabillities[4], with a FWHM of 30' for the 1.5 GHz observations (excluding the Lovell antenna) and a FWHM of 5.6' for the 5 GHz observations (including the Lovell antenna).

[4] Available at http://www.e-merlin.ac.uk/tech/

### 3.3 Brightness Temperature

Assuming an optically thick source, the expected observed brightness temperature for free–free emission from H II regions is of order 10 000 K. Electrons radiating via non–thermal mechanisms, however, lead to observed brightness temperatures that can be much higher (e.g., $10^{12}$ K); brightness temperatures could therefore be used to aid in source classification.

Assuming a Rayleigh–Jeans black–body curve, we calculate the brightness temperature, $T_{\rm B}$, for each detected source using:

$$T_{\rm B} \approx \frac{c^2}{2\nu^2 k_{\rm B}\Omega}\left(\frac{S_\nu}{10^{26}\,{\rm Jy\,W^{-1}\,m^2\,Hz}}\right) \qquad (2)$$

Where $k_{\rm B}$ is the Boltzmann constant, $c$ is the speed of light, $\nu$ is the central frequency of the combined 1.5 GHz observations (1518 MHz), $\Omega$ is solid angle that the synthesized beam subtends and $S_\nu$ is the measured flux density in Jy. The derived brightness temperature is only correct for sources that are resolved, i.e., when the synthesized beam is completely filled with the emitting source. As it stands, our brightness temperature for extended sources is likely to be an underestimate, due to us missing flux as a result of the lack of short baselines in these interferometric observations. The

**Table 4.** Angular Sizes of Extended and Resolved Sources at 1.5 GHz

| Source No | Major Axis ($''$) | Minor Axis ($''$) | Position angle ($°$) |
|---|---|---|---|
| 1* | $3.34^{+0.30}_{-0.63}$ | $1.85^{+0.26}_{-0.34}$ | $138.96^{+13.37}_{-10.99}$ |
| 2* | $1.86^{+0.12}_{-0.34}$ | $1.21^{+0.12}_{-0.27}$ | $14.93^{+17.18}_{-19.65}$ |
| 3* | $2.49 \pm 0.38$ | $0.96^{+0.14}_{-0.29}$ | $71.35^{+9.16}_{-6.76}$ |
| 4* | $1.46^{+0.09}_{-0.33}$ | $0.90^{+0.04}_{-0.17}$ | $9.51^{+11.49}_{-10.32}$ |
| 7* | $1.80^{+0.07}_{-0.21}$ | $1.25^{+0.06}_{-0.20}$ | $48.93^{+12.46}_{-11.37}$ |
| 8 | $0.694 \pm 0.094$ | $0.590 \pm 0.080$ | $123.65 \pm 33.87$ |

**Notes:** The superscript * in the Source No column indicates that the source is extended. The position angle of each source is measured anti-clockwise from the North.

brightness temperature observed for unresolved sources will also be an underestimate as the synthesized beam is larger than the true angular size of the source. Furthermore, our radio observations are probing optically thin conditions, resulting in a further reduction of the observed brightness temperatures. Therefore, the brightness temperatures derived both for all sources should be treated as lower limits. We present the peak brightness temperature for each detected source in Table 3 and display brightness temperature contours for the resolved sources in Figure 3.

### 3.4 Angular Sizes

We measure the angular size, position angle and sky co-ordinates of each extended source by fitting an ellipse to the $3\sigma_{\rm rms}$ contour. We optimize the fit by minimizing the $\chi^2$ statistic, assuming the error at each point on the contour is the same. We obtain errors in the fit via the $\Delta\chi^2$ method, i.e., by modifying each individual parameter to gain $\chi^2$ as a function of said parameter whilst fixing all other parameters to the best fit. The quoted $1\sigma$ errors are defined by the intersect where $\Delta\chi^2$ from the minimized $\chi^2$ is equal to 1.

For marginally resolved and unresolved sources, we fit 2–D Gaussian functions using the AIPS task JMFIT to determine the angular size, position angle and sky co-ordinates. The angular size properties for extended and resolved sources at 1.5 GHz are summarized in Table 4 and the sky co-ordinates are presented in Table 3.

### 3.5 Spectral Indices

Ideally, for unambiguous source classification, we require a combination of reliable spectral indices ($S_\nu \propto \nu^\alpha$), to gain a measure of the spectral energy distribution for each compact source, and maps made at multiple epochs with a long enough time baseline to measure SNR expansion ($\sim 10$ yrs: McDonald et al. 2001; Beswick et al. 2006; Fenech et al. 2008).

These current observations are the highest resolution maps of IC 10 yet, and hence, we have no previous observations to observe any expansion. Furthermore, to recover reliable spectral indices for resolved sources, we require both observations to have significantly overlapping *uv*-planes. For our current e–MERLIN observations however, this is not the case (see Figure 2) and we can only gain reliable spectral indices for sources that are unresolved at both 1.5 GHz and 5 GHz.

To find spectral indices for resolved sources, we make use of the spectral index maps provided in Heesen et al. (2011). These maps are taken at a much lower resolution than the presented e–MERLIN observations, and hence can only give us a general idea about the spectral index of the resolved sources combined with emission from their immediate surrounding environment.

### 3.6 Completeness & Reliability

We carried out Monte–Carlo simulations at 1.5 GHz to assess the completeness, $C$, and reliability, $R$, of our source detection method. At each band, we took a blank field from the widefield imaging (i.e., a noise map) and randomly added 400 Gaussian sources, each the same size as the naturally weighted synthesized beam. We assigned a flux density to each of these sources, drawn from a continous, uniform probability distribution between 2 and 12 times the rms noise level of the maps. We then used Aegean to recover the placed sources, setting a detection threshold at 3 times the map rms noise level. As in our analysis, only sources exceeding a threshold of $5\sigma$ are used to determine the statistical completeness of our source detection method. This process was repeated 1000 times to ensure a statically robust answer.

We split our data into flux bins, each of width 0.5 times the map rms noise. We define the completeness & reliability for each flux bin as:

$$C = \frac{N_{\rm M}}{N_{\rm P}} \qquad (3)$$

$$R = \frac{N_{\rm M}}{N_{\rm R}} \qquad (4)$$

Where $N_{\rm M}$ is the number of matched sources (a match is counted when the recovered source is spatially located within 1 synthesized beam of the placed source), $N_P$ is the number of placed sources and $N_{\rm R}$ is the total number of matched sources recovered by the source extractor that also within matched in flux (within $1\sigma_{\rm rms}$ uncertainty). The completeness is essentially a measure of the fraction of real sources our source extractor detects and the reliability is a measure of how often spurious detections may be mistaken for real sources.

The results from this analysis are summarised in Figure 5. We are complete down to approximately $7\sigma$ and reach 50% completeness at approximately $5\sigma$. The reliability plot tell us that our $5\sigma$ detections are very robust and it is unlikely that any of these sources are spurious detections. Modifying the detection threshold results in a translation of the completeness distribuion presented in the upper panel of Figure 5, where the chosen threshold broadly matches where the completeness reaches 50%. Changing the threshold to, for example $3\sigma$, increases the completeness at low flux levels, however, this does not result in an increase in the reliability, as additional sources may be spurious.

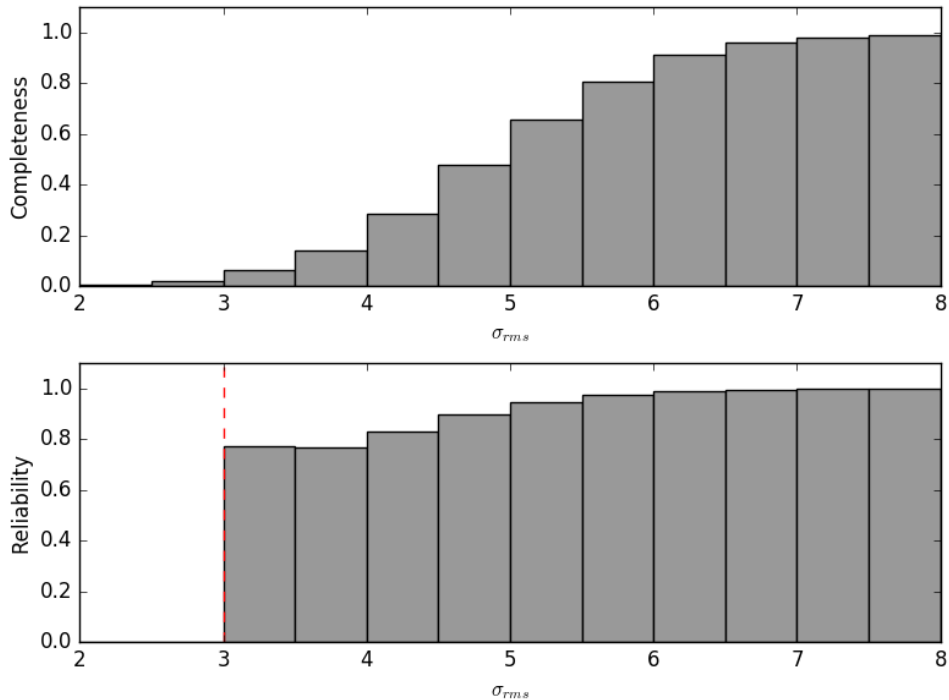

**Figure 5.** Completenesss and reliability plots for the combined 1.5 GHz observations of IC10. The uncertainty in all bins is <5%. In the lower plot, a vertical red dashed line represents the lower flux limit to which we can investigate the reliability.

## 4 DISCUSSION

### 4.1 Individual Source Notes

In this section we briefly discuss each source in turn, highlighting relevant information to aid in its classification.

**Source 1 (WBBH J002017+591839.7):** This is the brightest source that we detect in our e–MERLIN observations and it also has the largest angular size. There is evidence in the presented contour maps for a 'clean bowl', hinting that we are missing flux from this source due to the lack of short spacings in the e–MERLIN array. We find that this source is spatially coincident with emission peaks in the H$\alpha$, 70 µm maps and lower resolution 5 GHz and 1.5 GHz maps. This source also falls between 2 holes in the large-scale H I emission (see Figure 1). We derive a brightness temperature of approximatly 5000 K for this source, which is lower than expected. It is unlikely that this is due to a high optical thickness along the line of sight to the source, and is probably due to missing flux. The low resolution spectral index from the literature is flat ($\alpha \sim -0.3$). In view of the above, we classify this source as a compact H II region within IC 10.

**Source 2 (WBBH J002027+591728.7):** This source is spatially coincident with peaks of emission in the H$\alpha$ and 70 µm maps as well as lower resolution 5 GHz, and 1.5 GHz maps. It is also spatially coincident with a current major star forming region within IC 10 (Heesen et al. 2015), with numerous WR stars nearby. Moreover, this source is located within the highest peak of the H I distribution, indicating that SF is possibly being fuelled at this location. The low-resolution spectral index from the literature for this source is -0.1, which is consistant with a dominating thermal contribution in this region. The brightness temperature for this source is of order 8000 K, which is roughly what is expected for an H II region. Therefore, it is likely that this source is a compact H II region within IC 10.

**Source 3 (WBBH J002015+591853.9):** This source is particularly interesting. It has an unusual extended morphology, no counterpart in either the H$\alpha$ or 70 µm maps and does not appear to be associated with any diffuse H I emission. This source does not appear to be related to any star-forming regions within IC 10, but it is spatially coincident with a Chandra X–ray source. As this source also has a steep low–resolution spectral index ($\alpha \sim -1.0$), determined from lower resolution maps in the literature, it is likely to be a contaminating background galaxy. This source has a lower than expected brightness temperature of roughly 3000 K, which could be due to high optical thickness along the line of sight to this source or due to missing flux from short baselines. Future VLA A–Array 10 GHz observations would provide a reliable spectral index for this source and help determine whether it resides within IC 10.

**Source 4 (WBBH J002027+591706.1):** Like Source 2, this source is coincident with emission in the H$\alpha$ and 70 µm maps, the lower resolution 5 GHz, and 1.5 GHz maps as well as with a peak in the H I distribution. Source 4 is located on the outskirts of the Non–Thermal superbubble (see Section 4.5), and there are many WR stars in this general region of the galaxy. We determine a brightness temperature of roughly 5000 K, which is lower than expected and could be due to the same reasons as discussed for source 1. This source has a relatively flat low resolution spectral index ($\alpha \sim -0.4$) and is likely to be a compact H II region within IC 10.

**Source 5 (WBBH J002008+591540.2):** This source is unresolved and located towards the southern outskirt of IC 10's main disk. There are no corresponding sources at H$\alpha$ or 70 µm. As this source is unresolved, the derived brightness temperature must be treated as a lower limit and the true brightness temperature must be greater than 10000 K. If this source is of non-thermal origin, we expect that it must have an angular size much smaller than the synthesized beam. Furthermore, this source has a steep spectral index ($\alpha \sim -1.3$) as determined from our e–MERLIN observations and it is also coincident with a Chandra X–ray source. It is likely that this source is a background AGN.

**Source 6 (WBBH J002027+591843.7):** This source is unresolved and located towards the North-East of IC 10's main disk. Similarly to Source 5, there is no corresponding emission in either H$\alpha$ or 70 µm maps. We derive a lower limit for the brightness temperature of 8000 K, but this is likely to be due to the source filling a small fraction of the synthesized beam. This source is spatially located close to but separate from a Chandra X–ray source. It is surprising that we do not detect this bright source in the presented 5 GHz observations, resulting in an extremely steep spectral index estimate ($\alpha \leq -1.9$). From this evidence, it is likely that this source is a background source.

**Source 7 (WBBH J002018+591740.9):** This source has a resolved, diffuse structure with a sharp peak above the 5 $\sigma_{\rm rms}$ threshold. In Figure 3 we display a contour map that we convolved to a resolution of twice the naturally weighted synthesised beam to highlight this diffuse emission. We carried out the same analysis as for the other sources but with a beam that was twice the size. The peak brightness temperature for this source was found to be just 700 K, and this should be treated as a lower limit. This source is located roughly in the centre of IC 10's main disk, with corresponding peaks of emission in the H$\alpha$, 70 µm and the low resolution radio continuum maps. Moreover, this source has a flat low–resolution spectral index, suggesting that this source is possibly an H II region within IC 10 although more investigation will be required to fully classify this source. There is emission at a 3 $\sigma_{\rm rms}$ level located $\sim 2.5'$ towards the south–east of this source. If this emission is also associated with source 7, it is possible that this source is a young SNR and we are observing high brightness features in the expanding shell (see Section 4.4). Future, more sensitive, e–MERLIN 1.5 GHz and VLA 1.5 GHz A–Array observations will be required to determine if this source has an extended component. VLA 10 GHz A–Array observations could also be used to derive a resolved spectral index to aid in classifying this source.

**Source 8 (WBBH J002019+591802.3):** This source is partially resolved and located towards the centre of IC 10. This source is coincident with diffuse emission in the H$\alpha$ map and with peaks in the 70 µm and low resolution radio maps. We do not detect this source in the presented 5 GHz maps, resulting in a very steep inferred spectral index ($\alpha \leq -2.0$). We present a lower limit for the brightness temperature of order 8000 K. This source could either be a background source or a peak in a more extended H II region. Deeper e–MERLIN 5 GHz observations would be required to obtain a reliable spectral index for this source and to determine its membership to IC10.

**Source 9 (WBBH J002033+591942.9):** This source is another compact, unresolved source located towards the North–East of IC 10's main disk. The source appears to be extended to the East in Figure 3, but visual inspection suggests this to be a noise peak. This source appears isolated from IC 10's main disk, with no corresponding emission in either the H$\alpha$ or 70 µm maps. This source is faint with a lower limit to the brightness temperature of 4000 K, which is probably due to the source filling a small fraction of the synthesized beam. Due to its isolation, we classify this source as a background galaxy.

**Source 10 (WBBH J002008+591955.2):** This source is another unresolved source located towards the North–West of IC 10's main disk. This source looks like it could be associated with the northern ring in the H I distribution (see Figure 1) but it appears to be offset. There is no corresponding emission in either H$\alpha$ or 70 µm maps. We derive a lower limit to the brightness temperature of 3000 K which is possibly due to the source filling a small fraction of the synthesized beam. This source appears to be coincident with a low resolution XMM X–ray source, but due to the low resolution ($13''$, Wang et al. 2005) we cannot say with certainty that this source is the origin of the X–ray emission. This current evidence suggests that this source is a faint background source.

**Source 11 (WBBH J001951+591644.2):** This source is another faint, unresolved source that is located towards the West of IC 10's main disk. This source falls within extended emission in the 70 µm map but the H$\alpha$ map does not extend this far. This source is likely to be a background galaxy as it has a steep spectral index ($\alpha < -1.1$) and there are no coincident compact sources at any other wavelength. It could also be a peak in a more diffuse H II region but this is unlikely given the spectral index. The lower limit to the brightness temperature is still quite low ($\sim$2000 K) which is possibly due to the source filling a small fraction of the synthesized beam. Further investigation will be required to properly classify this source, but current evidence indicates it is likely to be a background galaxy.

### 4.2 Expected Background Source Count

We derive the expected number of background sources coincident with IC 10 by analysing source counts from the literature. We fit the source counts from Massardi et al. (2010) with a 4th order polynomial[5]. This was chosen as it is the lowest order polynomial that represents the data well. We then apply corrections for completeness (see Section 3.6) and end up with an estimate for the background count of $14.8 \pm 3.8$ sources in a circular field of radius 5'. It is likely that this estimate is a slight overestimate as any nearby extended sources will be resolved out, but this is expected to only affect brighter sources which contribute little to the overall counts. In a region of the same size centred on IC 10 (see Section 3.1) we detect 15 sources which is consistent with the hypothesis that all detected sources are background galaxies. As discussed in Section 4.1, it is clear from comparison with H$\alpha$ and 70 µm maps that at least 3 of our detections reside within IC 10, but it is obvious from this exercise that the majority of the detected sources are probably background galaxies.

---

[5] The raw data is available from http://web.oapd.inaf.it/rstools/srccnt/srccnt_tables.html.

**Table 5.** Determined H II Region properties

| Source | $N_{UV}$ | Stellar Cluster Mass | # SNe |
|---|---|---|---|
| 1 | $\geq 3.1 \times 10^{50}\,\mathrm{s}^{-1}$ | $\geq 22000\,\mathrm{M}_\odot$ | $\gtrsim 266$ |
| 2 | $\geq 1.3 \times 10^{50}\,\mathrm{s}^{-1}$ | $\geq 9500\,\mathrm{M}_\odot$ | $\gtrsim 115$ |
| 4 | $\geq 7.0 \times 10^{49}\,\mathrm{s}^{-1}$ | $\geq 5200\,\mathrm{M}_\odot$ | $\gtrsim 63$ |

**Notes:** # SNe is the total number of expected SNe over the entire lifetime of the H II region assuming a Miller–Scalo IMF with all stars greater than $8\,\mathrm{M}_\odot$ exploding as SNe.

### 4.3 Compact H II Regions Within IC 10

From our discussion in Section 4.1, we determined that sources 1, 2 and 4 are likely to be compact H II regions within IC 10. The source of the ionising Lyman continuum photons that produce these H II regions is the young massive O and B type stars (Strömgren 1939). If we assume that all detected radio emission from these sources at 1.5 GHz is thermal in origin, we can calculate the production rate of Lyman continuum photons required to produce the observed radio emission. By comparing these results with the outputs from stellar population synthesis models, we can determine an estimate for the total stellar mass contained within the star clusters responsible for the observed H II regions.

To estimate the total production rate of Lyman continuum photons for each detected compact H II region, we applied Equation 2 from Condon (1992), assuming an electron temperature of $10^4$ K. These estimates will be lower limits for the true Lyman continuum production rate as not only are we resolving out some of the extended emission associated with these soures, some Lyman continuum photons may also be absorbed by dust within the H II regions themselves. We then used Starburst99 (Leitherer et al. 1999) to produce models for the Lymann continuum production rate as a function of time for various stellar cluster masses. We assume a Miller–Scalo IMF (Miller & Scalo 1979), that all star formation takes places in a single instantaneous burst at t=0 and that all stars larger than $8\,\mathrm{M}_\odot$ explode as SNe. We also assume Geneva evolutionary tracks with standard mass loss at a metallicity of Z=0.004 (Charbonnel et al. 1993) as this metallicity closely reflects that observed for IC 10 (Garnett 1990; Leroy et al. 2006).

Each of the models for the Lyman continuum production rate have the same shape, the production rate is constant for the first $\sim 10^{6.5}$ yr before sharply declining due to the massive stars that produce the Lyman continuum emission exploding as SNe (see Figure 6). As we do not know the age of the observed H II regions, we do not know if any of the massive OB stars have already exploded as SNe. However, as these sources are also observable in the H$\alpha$ map (see Figure 1), it is likely that no SNe have yet disrupted the Lyman continuum production rate and that we are observing these H II regions in the early constant production phase. Comparing our observed Lyman continuum production rate with the constant phase predicted from Starburst99 will therefore give a reasonable estimate of the lower limit for the stellar cluster mass.

We determine the stellar masses for each detected H II region by interpolating between the constant production phase for each Starburst99 model. The derived lower limits for the cluster masses along with the total number of expected

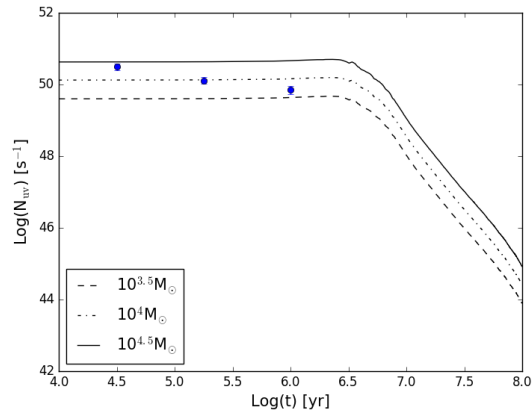

**Figure 6.** Starburst99 models of the Lyman continuum production rate as a function of time for 3 cluster masses. The solid line represents a $10^{4.5}\,\mathrm{M}_\odot$ cluster, the dot–dash line represents a $10^4\,\mathrm{M}_\odot$ cluster and the dashed line represents a $10^{3.5}\,\mathrm{M}_\odot$ cluster. We plot the lower limits for the Lyman continuum production rate for sources 1, 2 and 4 (blue dots). Note that we do not know the age of the observed H II regions and so the position on the time axis was arbitraily chosen to clearly identify where each source lies with respect to the constant Lyman continuum production phase.

SNe originating from these clusters are presented in Table 5. These estimates do not vary much with assumed metallicity, but do vary greatly with assumed IMF. Re-running the simulations again assuming solar metallicity results in a 5 percent increase in the cluster mass, however, re-running the simulations assuming a Salpeter IMF (Salpeter 1955) results in a 50 percent decrease in the cluster mass. The derived cluster mass for sources 1 and 2 are larger than the average stellar cluster found within the LMC whereas source 4 reflects a more typical cluster mass (Hunter et al. 2003). For reference, the derived cluster mass for source 1 is approximately half that of 30 Doradus in the LMC (Hunter et al. 1995; Andersen et al. 2009).

### 4.4 SNR Within IC 10

Although we cannot individually classify with confidence all detected sources within IC 10, we can use the source sizes and evolution timescales from the literature to constrain how many SNR we would expect to detect. It is useful to utilize observations of the LMC and SMC as these galaxies have similar environmental properties to IC 10 and are nearby, allowing for more complete studies of their SNR populations.

Studies of the SNR in the Magellanic clouds (see reference list in Filipović & Bozzetto 2016) reveal that the average size of an SNR is $\sim 40$ pc (Filipović & Bozzetto 2016), which is similar to the average size of SNR in M 33 (Long et al. 2010) and M 31 (Lee & Lee 2014). Scaling to our assumed distance of 0.7 Mpc, we would expect an average SNR to have an angular size of $\sim 12''$. e–MERLIN essentially begins to resolve out sources that have an angular size larger than $2''$ at 1.5 GHz[6]. A rough calculation using equations detailed in Draine (2011) reveals that SNR appearing to be $2''$ in size would be ~40 yr in age and still in the short–lived free expansion phase (Draine 2011). As most observed SNR are in the Sedov–Taylor phase of expansion (which remains observable for ~$10^4$ yr; Strong et al. 2007), even with a large sample of SNR, we would expect only a small fraction to be young SNR. It is therefore likely that we are resolving out most of the SNR within IC 10. We are however still sensitive to high brightness features within an expanding SNR shell and so it is possible that some of our observed sources are part of an expanding SNR shockfronts.

Through measurements of the integrated star–formation rate (SFR) of IC 10, we can gain an approximate estimate of the number of observable SNR that we would expect to find using high-resolution radio continuum observations. Following the calculations carried out in Condon (1992), assuming a Miller–Scalo IMF (Miller & Scalo 1979) with all stars with mass above $8\,M_\odot$ exploding as SNe and using the SFR from Hunter et al. (2012), we find the supernova rate is approximately $\nu_{SN} = 8.1 \times 10^{-4}\,\mathrm{yr}^{-1}$. With a typical lifetime of $10^4$ yr, one would expect to observe roughly 8 SNR. Brighter radio SNe (RSNe) are observable for a much shorter time period (~10 yr; Stockdale et al. 2001; Weiler et al. 2002) and so it is very unlikely that we will detect any RSNe in IC 10 at all.

There are other biases that will affect the detectability of SNR. For example synchrotron emission requires magnetic fields to produce emission. Magnetic fields are amplified in SNR shockfronts (Guo et al. 2012) and the base magnetic field strength is related to the density of the surrounding gas (Vallée 1995; Niklas & Beck 1997). As the gas density in dwarf galaxies is low, we expect a SNR within IC 10 to be intrinsically fainter than those taking place denser environments (such as M 82 and Arp 220).

As we expect to find so few SNR in general, it is unlikely that we are observing any young SNR with these e–MERLIN observations. Instead we are more likely to be observing compact H II regions and background sources. In general, e–MERLIN would be well suited to observe SNR at 1.5 GHz in galaxies at distances $> 4$ Mpc. At this distance, the average SNR would appear $\sim 2''$ in size and the SNR would not be resolved out due to the long baselines in the e–MERLIN array. For galaxies that are closer than 4 Mpc, e–MERLIN observations should be combined with VLA A–array observations which will provide the required sensitivity to large scale emission in order to identify possible SNR within these galaxies.

### 4.5 Non-Thermal Superbubble

The Non-Thermal Superbubble (NTSB) within IC 10 (Yang & Skillman 1993) is a remarkable example of SN feedback in a low density environment. The NTSB is adjacent to the largest current star forming site in IC 10 and at its centre resides one of the most massive stellar mass black holes ever discovered (IC 10 X-1, $\geq 23.1\,M_\odot$; Silverman & Filippenko 2008). There is diffuse X–ray emission in this area of the galaxy, which is similar to 30 Doradus in the LMC (Wang et al. 2005; Wang 1999). It has been speculated that this black hole is the remains of a Hypernova that is responsible for the NTSB (Lozinskaya & Moiseev 2007), although it has been argued that the superbubble could be the result of a collection of many ordinary SNe (Yang & Skillman 1993). The NTSB is dominated by emission of non-thermal origin, and the presence of an H I 'bubble' in this region presents evidence that there possibly exists a cavity in the ISM at this location (Heesen et al. 2015).

We do not see any compact radio sources within the NTSB in the e–MERLIN observations. We would not expect to see any SNe in this region due to the arguments given in Section 4.4, and this hypothesis can be further supported by considering the environmental conditions within the NTSB. The initial growth of superbubbles in general is thought to originate from the winds of massive stars which rarify their immediate environment (Krause et al. 2013). As these stars explode as SNe, their shockwaves will quickly propagate through the rarified ISM until they reach the outer, dense superbubble shell where they can more efficiently accelerate CRe and produce substantial radio emission. Hence we would expect to see a large radio 'shell' surrounding the NTSB where the SNe shockfronts essentially catch up with the edge of the superbubble.

It is interesting that we do not resolve a shockfront associated with the NTSB in these high–resolution observations. This could simply be due to the interferometer resolving out large scale emission (the NTSB itself is $\sim 50''$ across), however we would expect to see features in the shell that are smaller than $2''$. It is also possible that the expansion of the superbubble has slowed sufficently such that CRe cannot be effectively accelerated via the diffusive shock acceleration process in the expanding shell (Longair 2011). Recent estimates of the dynamical age and expansion velocity of the superbubble from Heesen et al. (2015) suggest that the latter is likely and that the NTSB could be in a 'fade–out' stage of its evolution.

### 4.6 Updated Integrated Flux Measurment

The radio–FIR relation is a very tight trend that holds over 4 orders of magnitude in luminosity (e.g. Yun et al. 2001), linking both radio continuum and FIR emission to massive star–formation. Dwarf Irregular galaxies are found to lie on this correlation, yet they are observed to have a deficiency in radio continuum when compared to the SFR derived from hybrid SF tracers (Kitchener et al. 2015). The reason for this deficency is interpreted to be due to CRe escaping from their low gas density environments (Klein et al. 1991; Lacki et al. 2010; Kitchener et al. 2015), which needs to be coupled with escaping far–UV photons to explain their location on the radio–FIR relation (the radio–FIR conspiracy: Lacki et al. 2010).

Total flux measurements of galaxies are typically determined through single dish observations, which are often contaminated by background sources. As dwarf irregular galaxies are intrinsically faint in radio continuum, the background sources could dominate the observed radio emission, resulting in an incorrect measurement of the total radio emission from the galaxy itself. With this current study, we identify potential background sources and use our high–resolution

---

[6] See e–MERLIN Cycle 4 capabilities avaliable at http://www.e-merlin.ac.uk/observe/cycle4.html

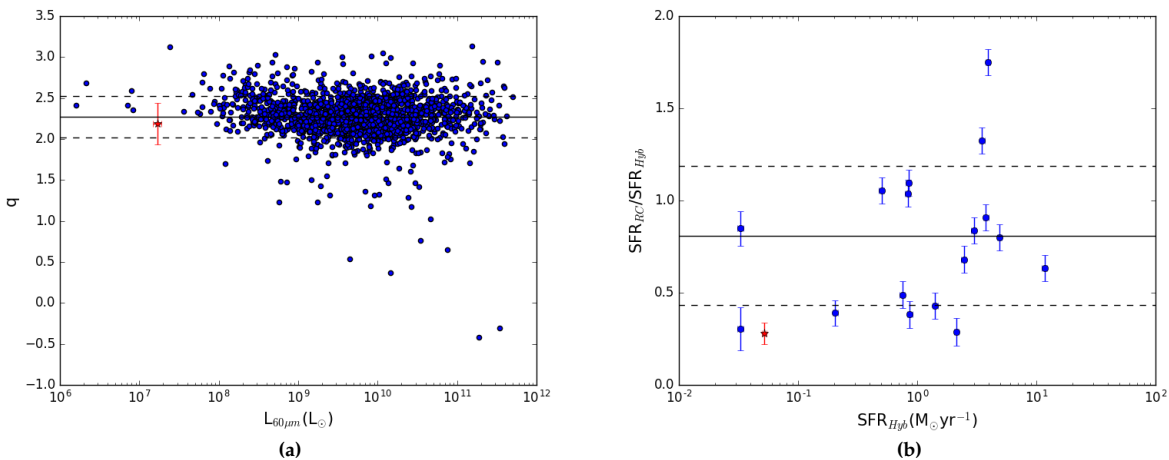

**Figure 7.** (a) Distribution of q–values plotted as a function of IRAS 60 μm luminosity. The blue points are taken from Yun et al. (2001) and the red star presents IC 10's location. The solid line represents the mean q–value for the sample and the dotted lines present the 1 σ devations from the mean. (b) Ratio of the Radio–Continuum SFR determined via the Condon relation (Condon 1992) to the Hybrid SFR as a function of Hybrid SFR for the Heesen et al. (2014) sample. The red star presentes IC 10's location. The solid line represents the mean of the sample excluding dIrr galaxies and the dotted lines represent 1 σ deviations from the mean.

maps to update the integrated flux density measurements in the literature accordingly.

We use the measurements from Chyży et al. (2016), which are $377 \pm 11$ mJy at 1.43 GHz and $222 \pm 9$ mJy at 4.85 GHz, as the integrated flux measurements for IC 10. These measurements have already been corrected for the presence of 4 background sources that are observed in the halo of the galaxy, but not for the existence of background sources within the main star–forming disk. We use our high resolution e–MERLIN maps to identify background sources within IC 10's main disk, but measure the total flux density for these sources from lower resolution VLA maps presented in Heesen et al. (2015). The e–MERLIN sources can easily be identified in the lower resolution VLA maps and by using this method, we minimise the amount of flux that we miss due to missing short spacings.

As the VLA maps were taken at 1.5 GHz and 6 GHz, we use the spectral index between the Chyży et al. (2016) measurements ($\alpha = -0.43 \pm 0.09$) to find the expected integrated flux densities at these frequencies. We then measure the flux density of sources that are determined to be background sources from the VLA maps using the AIPS task JM-FIT. We subsequently subtract these sources (sources 3, 5, 6, 9 and 10), along with 4 additional faint sources detected in IC 10's halo to obtain a corrected flux density of $352 \pm 11$ mJy at 1.5 GHz and $199 \pm 9$ mJy at 6 GHz.

Subtracting the contaminating background sources in IC 10's main disk results in only a 4 per cent decrease in the flux density at 1.5 GHz and a 2 per cent decrease at 6 GHz. Contaminating background sources do not contribute significantly to the overall flux density measured from IC 10 at radio wavelengths and previous analysis concerning its integrated properties remain valid.

We combine this updated radio flux density with IRAS 60 μm and 100 μm flux densities taken from Fullmer & Londsale (1995) to find $q_{IC10} = 2.19 \pm 0.25$. The q–parameter is essentially a measure of the radio–FIR relation and is defined in Condon et al. (1991) as:

$$q = \log\left(\frac{\text{FIR}}{3.75 \times 10^{12}\,\text{W m}^{-2}}\right) - \log\left(\frac{S_{1.4\text{GHz}}}{\text{W m}^{-2}\,\text{Hz}^{-1}}\right) \quad (5)$$

$$\text{FIR} = 1.26 \times 10^{-14}\,(2.58 S_{60\,\mu\text{m}} + S_{100\,\mu\text{m}})\,\text{W m}^{-2} \quad (6)$$

Where $S_{60\,\mu\text{m}}$ and $S_{100\,\mu\text{m}}$ are the IRAS 60 μm and 100 μm flux densities measured in Jy. It is clear from Figure 7a that IC 10 lies within 1 σ of the radio–FIR relation found for more massive star forming galaxies (Yun et al. 2001). This result agrees with earlier authors who find that low luminosity dwarf galaxies continue to follow the radio–FIR relation (Chyży et al. 2011; Roychowdhury & Chengalur 2012).

We also transform the updated radio flux measurement to a SFR via the Condon relation (Condon 1992) and compare this with the SFR measured from hybrid SF tracers ($0.052 \pm 0.003\,\text{M}_\odot\,\text{yr}^{-1}$; Heesen, private communication) to find that IC 10 lies just below the radio–SFR relation from Heesen et al. (2014, see Figure 7b). We would expect IC 10 to be radio deficient due to the escape of CRe in a galactic wind (Chyży et al. 2016), yet it is interesting to note that this doesn't result in a significant deviation from the radio–SFR relation. We cannot currently say whether IC 10 is an example of the radio–FIR 'conspiracy' as the sample size used to determine the radio–SFR relation is small with a large scatter. Future studies with a larger sample would reveal the true significance of the deviation.

## 5 CONCLUSIONS

We present high-resolution e–MERLIN observations of the nearby Dwarf Irregular galaxy IC 10 as a part of the LeM-MINGs project. Our main conclusions are as follows:

(i) Our e-MERLIN observations reveal 11 compact radio sources at 1.5 GHz coincident with IC 10's star-forming disk, 5 of which also have detections at 5 GHz. Although we cannot conclusively classify all sources on the basis of their spectral index due to differences in sampling the uv–plane, we gain insight into the nature of each source through the use of low–resolution multi–wavelength data, the positional coincidence with features at other wavelengths such as H$\alpha$ and 70 µm, and an estimate of their brightness temperature.

(ii) Based on our analysis of the expected background source counts, the number of detected sources is consistant with them all being background sources. However, as some sources closely trace compact emission in maps at other wavelengths, giving us confidence that some of these sources do in fact belong to IC 10. We find that of the 11 sources, 3 likely reside in IC 10 (sources 1,2,4), all of which are HII regions, 5 sources are background sources (sources 3,5,6,9,10) and 3 are yet to be conclusivly classified (sources 7,8,11).

(iii) Based on the timescale over which SNR are typically observable and the resolution limitations of these high–resolution interferometry observations, we expect that we will not be able to detect any SNR with the present e–MERLIN observations. The only SNe we would expect to be able to detect are very young RSNe, but it is unlikely that we would observe any due to a combination their short observable timescales and low SFR observed in IC 10. We explore other possible factors that could affect SNR detectability in dwarf Irregular galaxies and conclude that e–MERLIN alone is well suited to study SNR in galaxies located at distances greater than 4 Mpc due to the spatial frequencies covered by the array. For galaxies closer than 4 Mpc, additional VLA observations are required to fully cover the spatial scales that SNR candidates occupy.

(iv) The same reasons as mentioned above explain why we see no compact sources within the NTSB in IC 10 and that the detected sources that belong to IC 10 are likely to be compact HII regions. Based on Starburst99 models, we determine lower limits for the stellar cluster masses required to produce these observed compact HII regions; the largest one is about half the size of 30 Doradus in the LMC.

(v) We present updated integrated flux densities for IC 10, where we subtract the contribution from background sources from literature data. We find a 4 per cent reduction in the measured flux at 1.5 GHz and a 2 per cent reduction at 4.85 GHz.

(vi) With this updated measurement, we find that IC 10 lies below the radio–SFR relation derived in (Heesen et al. 2014), yet still remains on the radio–FIR relation. Further studies with larger samples are required to place significance on IC 10's radio deficiency to determine whether it is an example of the radio–FIR 'conspiracy'.

Further observations are required to fully classify the observed resolved sources. Ideally, VLA A–array observations taken at 10 GHz will match the *uv*–coverage of the current 1.5 GHz observations and allow for reliable spectral indices for all detected sources to be determined and further A–Array observations at 1.5 GHz will aid in the detection of SNR.

Regarding future LeMMINGs deep targets, 10 GHz A–Array VLA observations for all targets would greatly aid in classifying detected compact sources, whereas 1.5 GHz A–Array VLA observations would only be necessary for targets that are located closer than $\sim$4 Mpc to improve sensitivity to more extended emission.

Because of the low integrated SFR of individual dIrr galaxies, observations suffer from low-number statistics. In order to make progress, rather than targeting individual dIrr systems, an ensemble would need to be observed and statements made on a statistical basis. In contrast, larger star–forming spiral and starburst galaxies will contain many more compact sources and could be analysed on a galaxy by galaxy basis; it is suggested the next phase of LeMMING should place its focus there.

## ACKNOWLEDGEMENTS

JW acknowledges support from the UK's Science and Technology Facilities Council [grant number ST/M503514/1].

EB acknowledges support from the UK Science and Technology Facilities Council [grant number ST/M001008/1].

VH acknowledges support from the UK Science and Technology Facilities Council [grant number ST/J001600/1].

e–MERLIN is a National Facility operated by the University of Manchester at Jodrell Bank Observatory on behalf of STFC.

ParselTongue was developed in the context of the ALBUS project, which has benefited from research funding from the European Community's sixth Framework Programme under RadioNet R113CT 2003 5058187.

This paper has been typeset from a T<sub>E</sub>X/L<sup>A</sup>T<sub>E</sub>X file prepared by the author.